\documentstyle[12pt]{article}

\newcommand{\lsim}{\mbox{ $
{}_{{}_{\textstyle\sim}}  \! \! \! \! \! {}^{{}_{\textstyle<}}$}}
\begin{document}
\renewcommand{\thefootnote}{\fnsymbol{footnote}}
\thispagestyle{empty}
\vspace*{-1 cm}
\hspace*{\fill}  \mbox{WUE-ITP-95-003} \\
\vspace*{1 cm}
\begin{center}
{\Large \bf Mass Bounds for the Neutral Higgs Bosons in
the Next--To--Minimal Supersymmetric Standard Model
\\ [3 ex] }
{\large F.~Franke\footnote{Supported by Cusanuswerk \\ \hspace*{0.5cm}
email: fabian@physik.uni-wuerzburg.de}, H. Fraas
\\ [2 ex]
Institut f\"ur Theoretische Physik, Universit\"at W\"urzburg \\
D-97074 W\"urzburg, Germany}
\end{center}
\vfill

{\bf Abstract}

In the Next--To--Minimal Supersymmetric Standard Model (NMSSM), the
Higgs and
neutralino/chargino sectors are strongly correlated by four
common parameters at tree level.
Therefore we analyze the experimental data from
both the search for Higgs bosons as well as for neutralinos and charginos at
LEP 100 in order to constrain the parameter space and the
masses of the neutral Higgs
particles in the NMSSM. We find that
small singlet vacuum expectation values
are ruled out, but
a massless neutral Higgs scalar and pseudoscalar is not excluded
for most of the parameter space of the NMSSM.
Improved limits from the neutralino/chargino search at LEP 200,
however, may lead
to nonvanishing lower Higgs mass bounds.
\vfill
\begin{center}
March 1995
\end{center}
\newpage
\setcounter{page}{1}
\section{Introduction}
The search for Higgs bosons is one of the most exciting challenges at
the present and future high energy colliders.
Suppose that a Higgs particle will be discovered then the next question
to be answered is whether it belongs to the
Standard Model (SM) or to a model with an enlarged Higgs sector.

Supersymmetric models are the most attractive candidates for such
extended models.
The Minimal Supersymmetric Standard
Model (MSSM), which technically solves the hierarchy problem of the SM,
contains two Higgs doublets with vacuum expectation values $v_1$ and
$v_2$ ($\tan\beta=v_2/v_1$). So the Higgs sector consists of
five physical Higgs bosons,
two CP-even and one CP-odd neutral scalars, and a pair of
charged scalars. There is only a weak connection with
the neutralino/chargino sector by one common parameter
$\tan\beta$.

In the context of superstring and GUT theories \cite{barr,nilles,derendinger}
often
the simplest extension of the MSSM by a Higgs singlet superfield is
favored. This
Next-To-Minimal Supersymmetric Standard Model (NMSSM)
may also
provide a natural solution for the $\mu$-problem of the MSSM
\cite{mu}. It contains five physical neutral
Higgs bosons, three Higgs scalars $S_a$ ($a=1,2,3$)
and two pseudoscalars $P_b$ ($b=1,2$) \cite{higgs,drees}.
As a further crucial difference to the MSSM,
the Higgs- and neutralino/chargino sectors of the NMSSM are strongly
correlated: Once the parameters of the Higgs sector are fixed,
also the masses and mixings of the neutralinos and charginos are
determined by only one further parameter,
the gaugino mass $M$.

For both the SM and MSSM lower limits for the Higgs masses have been derived
from the LEP experiments.
For a SM Higgs boson, there exists a lower mass bound of 63.5 GeV
\cite{alephsm}.
In the MSSM, the situation is more complex due to the larger particle content.
Here the lower mass bound is 44 GeV
for the lightest scalar Higgs and 21 GeV for the pseudoscalar Higgs
particle \cite{alephmssm}.

The purpose of this letter is to derive Higgs mass bounds and to constrain
the parameter space of
the NMSSM by the results of the LEP experiments.
Due to the strong correlation between the Higgs and neutralino sectors,
constraints originate not only from the unsuccessful
direct Higgs search via
$e^+e^- \longrightarrow ZS_a, \,
S_aP_b$ and the contribution
$Z \longrightarrow S_aP_b$
to the total $Z$-width, but also from the unsuccessful neutralino and
chargino search. We therefore make use of the LEP constraints
for the neutralino
sector of the NMSSM previously obtained in ref.~\cite{franke}.

In our analysis the one-loop radiative corrections
of the Higgs masses due to stop and top loops are included
\cite{elliott}. With respect to CPU time
we did not take into account corrections from Higgs and
Higgssino loops which, however, are not expected to change our
results significantly.

The Higgs mass spectrum of the NMSSM has been the topic of several previous
studies.
Kim et al.~\cite{kimhiggs} analyzed the masses of the scalar and
pseudoscalar Higgs particles at tree level
and obtained excluded regions for certain of the NMSSM parameters by
assuming a rough discovery limit of 10 fb for the
process $e^+e^- \longrightarrow b\bar{b}S_1$.
Elliott et al.~\cite{elliott} computed the radiative corrections and
presented the scalar and pseudoscalar masses in dependence of the
mass of the charged Higgs for some parameters. But both authors
did not address the question in detail which parameters and masses
are already excluded by the present experimental results from LEP.
Ellwanger et al.~\cite{ellwanger} studied the particle spectrum of
the NMSSM under the assumption of universal soft symmetry breaking
terms at the GUT scale. They
included the LEP bound for the $ZZS_a$ coupling, but did not
discuss neither the parameter dependence of the Higgs masses
nor the constraints of the parameter space.

This letter is organized as follows: In Section 2 we briefly
describe the Higgs-Sector of the NMSSM. With the experimental constraints
described in Sec.3, we analyze in Sec.4 the restrictions for the
parameter space of the NMSSM and discuss the resulting Higgs mass bounds
and their dependence on the model parameters.

\section{The Higgs sector of the NMSSM}
The superpotential of the NMSSM is given by
\begin{eqnarray}
W & = & \lambda \varepsilon_{ij} H_1^i H_2^j N -\frac{1}{3} k N^3
\nonumber \\ & &
+h_U \varepsilon_{ij} \tilde{Q}^i \tilde{U} H_2^j
-h_D \varepsilon_{ij} \tilde{Q}^i \tilde{D} H_1^j
-h_E \varepsilon_{ij} \tilde{L}^i \tilde{R} H_1^j \; ,
\end{eqnarray}
where $H_1 = (H_1^0,H^-)$ and $H_2=(H^+,H_2^0)$ are the $SU(2)$ Higgs
doublets with hypercharge $-1/2$ and $1/2$,
respectively, $N$ is the Higgs singlet with hypercharge
0, and $\varepsilon_{ij}$ is totally antisymmetric with
$\varepsilon_{12}=-\varepsilon_{21}=1$. The notation of
the squark/slepton doublets and singlets is conventional,
generation indices are understood.
In the following we neglect all quark and lepton couplings apart
from that of the top quark.

The superpotential leads to the tree-level Higgs potential
at the supersymmetric mass scale
\begin{eqnarray}
V_{\mbox{Higgs}} & = &
\frac{g^2+g^{'2}}{8}( |H_1|^4 + |H_2|^4 )
- \frac{g^2+g^{'2}}{4}|H_1|^2|H_2|^2
\nonumber \\ & &
+\frac{g^2}{2}|H_1^{i \ast}H_2^i|^2
+ \lambda^2 [ ( |H_1|^2+|H_2|^2)|N|^2
+|\varepsilon_{ij}H_1^iH_2^j|^2 ]
\nonumber \\ & &
+k^2 |N|^4
-\lambda k (\varepsilon_{ij}H_1^iH_2^jN + \mbox{h.c.})
+ V_{\mbox{soft}}
\; ,
\end{eqnarray}
where $g'$, $g$ are the usual $U(1)$ and $SU(2)$ gauge couplings, and
the soft symmetry breaking potential is
\begin{eqnarray}
V_{\mbox{soft}} & = &
m_1^2 |H_1|^2 + m_2^2 |H_2|^2+m_3^2 |N|^2
\nonumber \\ & &
+m_Q^2 |\tilde{Q}|^2 + m_U^2 |\tilde{U}|^2 + m_D^2 |
\tilde{D}|^2
\nonumber \\ & &
+m_L^2 |\tilde{L}|^2 + m_E^2 |\tilde{R}|^2
\nonumber \\  & &
-(\lambda A_\lambda \varepsilon_{ij} H_1^i H_2^j N + \mbox{h.c.})
-(\frac{1}{3}kA_k N^3 + \mbox{h.c.})
\nonumber \\  & &
+(h_t A_t \varepsilon_{ij} \tilde{Q}^i \tilde{U} H_2^j
+\mbox{h.c.})  \; .
\end{eqnarray}

After eliminating the "soft" Higgs masses
$m_1$, $m_2$, $m_3$ by minimizing the Higgs potential with
respect to the vacuum expectation values
the Higgs sector can be
parametrized in terms of
six free parameters: the
ratio $\tan\beta=v_2/v_1$
of the vacuum expectation values of the Higgs doublets,
the vacuum expectation value $x$ of the Higgs
singlet, the trilinear couplings in the superpotential
$\lambda$ and $k$ and the "soft" scalar masses
$A_\lambda$ and $A_k$.
In addition, the radiative corrections to the scalar
potential due to top and stop loops
\cite{elliott} depend on
the "soft" top mass $A_t$ and the mass eigenvalues of the
scalar top quarks
$m_{\tilde{t}_1}$ and $m_{\tilde{t}_2}$, so that in total
there are nine parameters which determine the masses and
couplings of the five neutral Higgs bosons.

Decomposing the Higgs fields into their real and imaginary parts,
the $3\times 3$ mass matrices for the scalar and pseudoscalar
Higgses decouple with one of the CP-odd eigenstates being a
massless Goldstone boson.

Due to theoretical considerations we restrict the parameter range
as follows \cite{higgs,elliott,pietroni}:
\begin{enumerate}
\item As a sufficient condition for no explicit
CP violation in the scalar sector we choose
$\lambda$, $k$, $A_\lambda$, $A_k$ to be real and positive.
\item The vacuum state can be chosen such that
$v_1,v_2,x > 0$.
\item Assuming that perturbative physics remains valid up to the
unification scale we
bound the couplings $\lambda$ and $k$ by
their fixed-points
\begin{equation}
\label{fixed}
\lambda \leq 0.87 \, , \hspace{1cm} k \leq 0.63 \;  .
\end{equation}
\item The eigenvalues of the Higgs mass squared matrices are positive.
For the charged Higgs masses this condition is equivalent to the condition
that the vacuum does not break QED in the charged Higgs sector.
\item There exists no alternative lower minimum of the Higgs potential with
vanishing vacuum expectation values.
\end{enumerate}

For the analysis of the experimental constraints the couplings of the
scalar Higgs to two $Z$ bosons and to a
pseudoscalar Higgs and a
$Z$ boson are crucial. They are displayed in Fig.~1 where
$U^S$ and $U^P$ are the transformation matrices between the interaction
and mass eigenstates of the CP-even and CP-odd Higgs bosons, respectively:
\begin{eqnarray}
\label{higgs}
\left( \begin{array}{c} S_1 \\ S_2 \\ S_3 \end{array} \right) & = &
\frac{1}{\sqrt{2}} {U^S}
\left( \begin{array}{c} \mbox{Re} H_1^0 \\
\mbox{Re} H_2^0 \\ \mbox{Re} N \end{array} \right) -
\left( \begin{array}{c} v_1 \\ v_2 \\ x \end{array} \right) \; ,
\\ & & \nonumber \\
\left( \begin{array}{c} P_1 \\ P_2   \end{array} \right) & = &
\frac{1}{\sqrt{2}}
U^P
\left( \begin{array}{c} \mbox{Im} H_1^0 \\
\mbox{Im} H_2^0 \\ \mbox{Im} N \end{array} \right) \; .
\end{eqnarray}

\section{Experimental constraints}
Constraints of the NMSSM parameter space arise from searches for
both Higgs bosons and neutralinos at LEP.
Due to the unsuccessful direct and indirect Higgs search
the masses of the Higgs particles and the parameter space of the NMSSM
are constrained by
\begin{enumerate}
\item the limit for the factor $\xi^2_a=(U^S_{a1}\cos\beta+
U^S_{a2}\sin\beta )^2$ by which the production rate for
$e^+e^- \longrightarrow ZS_a$ ($a=1,2,3$)
in the NMSSM is reduced compared to the SM.
The ALEPH collaboration performed a detailed analysis of the
lower Higgs mass bound compatible with a model-dependent $\xi^2$
\cite{alephmssm} .
The 95 \% c.l. limits can be found in Fig.~2. Here one has to
distinguish between three different scenarios, where $S_a$
decays invisibly into two LSP's or
into two pseudoscalars with masses smaller than $2m_b$
or like a SM-Higgs, respectively.

\item limits from the direct search for the pseudoscalar Higgs in the processes
\begin{eqnarray}
\label{4b}
e^+e^- \longrightarrow S_1P_1 & \longrightarrow & b\bar{b}b\bar{b},
\: \tau\bar{\tau}\tau\bar{\tau},   \: b\bar{b},\: \tau\bar{\tau}.
\end{eqnarray}
They are given by the L3 collaboration as a function of the masses of the
scalar and pseudoscalar Higgs bosons
\cite{l3old}.
\item the upper limit \cite{l3old}
\begin{equation}
\Delta \Gamma_Z \leq 35.1 \; \mbox{MeV}
\end{equation}
of Higgs physics contributing to the total Z-width via
\begin{equation}
Z\longrightarrow S_aP_b \; (a=1,2,3; \; b=1,2)  .
\label{zsp}
\end{equation}
\end{enumerate}
The relevant cross sections and decay rates can
be easily derived
with the Feynman rules in Fig.~1.

The constraints from
direct and indirect neutralino search at LEP were presented
in \cite{franke} and included in our analysis
as far as they also constrain the
parameters of the Higgs sector.
\section{Higgs mass bounds in the NMSSM}
{}From the experimental limits of Sec.~3 we extract mass bounds for
the neutral Higgs bosons in the following way: We scan for fixed values of
$x$, $\lambda$, $k$, $\tan\beta$
as well as $A_t$ and the stop masses
the parameters $A_\lambda$ and $A_k$ over the theoretically allowed range.
For each set of parameters
the masses and mixings of the neutral Higgs bosons,
the $\xi^2$ factors and the decay rates (\ref{4b}) and
(\ref{zsp}) are computed and compared to
the LEP results in order to find
the lower mass bound.
Since in addition to the nine parameters of the Higgs sector
the gaugino mass parameter $M$ is sufficient to determine
also the masses and couplings of the neutralinos
we have to respect
constraints of the
parameter space from neutralino search illustrated in Fig.~3.

Fig.~2 shows a typical example for a scenario with a light LSP of
about 10 GeV, which is well below the
current lower bound for a MSSM neutralino.
Depicted is the limit for $\xi^2$ from LEP and the theoretical range
as a function of the mass of the lightest scalar Higgs $S_1$.
The lower mass bound for $S_1$ lies between 37 GeV (SM decay)
and 43 GeV (invisible decay). For the large singlet vacuum expectation
value $x=1000$ GeV
this bound is not affected by the limits (\ref{4b}) and
(\ref{zsp}) because in this case the lighter pseudoscalar Higgs
is almost a pure singlet.
Since for $m_{S_1} > 20$ GeV the invisible decay into
two neutralinos is kinematically possible the
dominant decay channel
depends on the choice for $A_\lambda$ and $A_k$.

For such scenarios  with a light, singlet-like neutralino a
very light neutral scalar Higgs of some GeV is excluded by the present
LEP data whereas for the pseudoscalar Higgs no such restriction exists.

The constraints for the parameters $A_\lambda$ and $A_k$ and the
Higgs masses for fixed $\lambda=0.4$, $k=0.02$, $\tan\beta=2$ and
various values of the singlet vacuum
expectation value $x$, the stop masses and $A_t$ are studied in
Figs.~4 and 5.
The $\lambda$-dependence of the lower bound of the scalar Higgs boson mass
is shown in Fig.~6 for $|M|=400$ GeV, $\tan\beta=2$ and two values
$x=100$ GeV and $x=1000$ GeV of the singlet vacuum expectation value.
The results do not substantially change for other choices of
$\tan\beta$ as long as the bottom contributions to the Higgs mixing
matrix can be neglected.

In order to combine all
LEP constraints from neutralino and Higgs search
we first present in Fig.~3
the excluded parameter space in the $M-x$ plane from unsuccessful
neutralino search at LEP (for details see
\cite{franke}). Fig.~3 shows that for
$\lambda=0.4$, $k=0.02$, $\tan\beta=2$ all $x$ values are allowed.
Contrary to this scenario, however,
our analysis in \cite{franke}
resulted in
lower $x$ bounds of about $50-100$ GeV
for most values of $\lambda$, $k$, and $\tan\beta$.
This gap of small allowed $x$ values
is expected to be closed by
LEP 200 unless a chargino will be found. But as we will show in
the following, combination of all present experimental constraints
from the gaugino/higgsino and Higgs sectors already leads to the exclusion of
very small $x$ values ($x \lsim 14$ GeV for $\tan\beta=2$).

In Fig.~4 the excluded
$A_\lambda -A_k$ region and
the allowed Higgs masses are shown for three values of $x$ with
$A_t=0$ GeV and the stop masses $m_{\tilde{t}_1}=150$ GeV and
$m_{\tilde{t}_2}=500$ GeV fixed, whereas
in Fig.~5 these parameters are varied with $x=1000$ GeV.
In the $A_\lambda -A_k$ plane (Fig.~4 (a), (c), (e)),
also plotted are the contour lines for the
masses of the Higgs bosons. There, the region above the
$m_{S_1}=0$ contour line is forbidden because the
mass squared would become negative. The domain beyond the dashed line
is excluded since there exists an alternative lower minimum of the Higgs
potential with vanishing vacuum expectation values.
The shaded region is forbidden
due to the experimental constraints given in Sec.~3. The allowed
mass region is shown in the
$m_{S_1}-m_{P_1}$ plane (Fig.~4 (b), (d), (f)), where
the solid line encloses the theoretical Higgs mass spectrum.

The dependence of the allowed range of parameters and Higgs masses on
the stop masses is rather weak. Larger stop masses as in
Fig.~5 (a) and (b) only lead to an insignificant increase of
the allowed domain.
On the other hand, the mass bounds are very sensitive
to the choice of the parameter $A_t$.
As shown in Fig.~5 (c) and (d),
large $A_t$ values
combined with
relatively small stop masses may substantially restrict the parameter space
and therefore cause very strong mass bounds.

Before continuing the discussion of the experimental constraints
let us shortly describe the theoretical Higgs mass range for
different parameters $x$, $\lambda$ and $k$. Generally, one
always finds values for $A_\lambda$ and $A_k$ leading to a
massless scalar Higgs boson. But, as Fig.~4 shows, for the lightest
pseudoscalar Higgs particle the lower mass bound
increases with increasing values for $x$, $\lambda$ and $k$.
Also the upper mass bounds depend on the parameters $x$, $\lambda$, $k$
in the same way: both the
scalar and pseudoscalar Higgs bosons
can be heavier for
larger values of $x$, $\lambda$ and $k$.

For large $x$ values ($x \gg m_Z$) the light pseudoscalar
becomes approximately a pure singlet
\cite{higgs} so that
there are only weak restrictions on the pseudoscalar Higgs mass
by experimental results
in Fig.~4 (b) and (d).

Since for smaller $x$ values the singlet components of the
lightest scalar and pseudoscalar Higgs bosons decrease,
the lower mass bounds approach the respective values of the
MSSM. The smaller the singlet vacuum expectation value $x$
becomes the larger the coupling $\lambda$ must be in order to
respect the constraints from neutralino search and to
allow for heavy Higgs bosons $S_1$ and $P_1$ which are not excluded
despite their MSSM-like mixing type. Keeping the maximal
$\lambda$ value restricted as suggested by eq.~(\ref{fixed}),
the upper mass bound for the lightest scalar becomes so small
that values $x \lsim 14$ GeV are excluded for $\tan\beta=2$.
For $x \geq 14 $ GeV, however, very light Higgs particles are
generally allowed provided the other parameters are properly
chosen.

Fig.~6 combines all constraints from neutralino and
Higgs search and shows the dependence of the lower mass bounds for
the scalar Higgs boson on the coupling $\lambda$. For this plot
with the singlet vacuum expectation values $x=100$ GeV and $x=1000$ GeV
we fixed the gaugino mass parameter at $|M|=400$ GeV. This value
minimizes the neutralino constraints but leads to a gluino mass
not much larger than 1 TeV as suggested by naturalness arguments,
by the hierarchy
problem \cite{hier}, and by fine tuning constraints \cite{fine}.

The graphs in Fig.~6 start with the minimal
$\lambda$ value
compatible with the constraints from the neutralino sector
($\lambda_{\mbox{min}}\approx 0.04$ for $x=1000$ GeV and
$\lambda_{\mbox{min}}\approx 0.35$ for $x=100$ GeV).
Within a small region of this starting point, the
lower mass bounds vanish, but finally rise sharply.
For smaller $x$ values, the curves move to larger couplings
$\lambda$ until for $x \lsim 14$ GeV the whole parameter space
is excluded due to the bound $\lambda \leq 0.87$ of eq.~(\ref{fixed}).

Improved constraints in the gaugino/higgsino sector by LEP 200 will possibly
raise the minimal allowed $\lambda$ value for a given
parameter $x$. According to the results of Fig.~6 this may cause
nonvanishing lower bounds for the mass
of the scalar Higgs boson.
Suppose that LEP 200 sets a lower chargino mass bound of 80 GeV.
Then for $x=100$ GeV and $\tan\beta=2$ the minimal $\lambda$ value becomes
about 0.7 leading to a mass bound for the scalar Higgs of
approximately 40 GeV even if improved constraints from Higgs search are
not considered. Also the lower bound for the singlet vacuum expectation
value $x$ may be raised up to 80 GeV.
That makes clear how in the NMSSM
Higgs constraints may be combined with
limits from unsuccessful neutralino search in order to
constrain the parameter space very effectively and obtain
mass bounds.

So the experimental constraints from the unsuccessful Higgs and
neutralino search at
LEP 100 lead to significant restrictions of the parameter space of the
NMSSM especially in scenarios with a light neutralino but are not yet
powerful enough to set general lower bounds for the neutral
scalar or pseudoscalar Higgs masses. Small singlet vacuum
expectation values, however, are already ruled out.
When the bounds from chargino and neutralino search will be improved
by LEP 200, it may be possible to exclude $x$-values up
to about 80 GeV and to
impose mass bounds for the neutral Higgs bosons at least for some
further singlet vacuum expectation values.

\newpage
\section*{Figure Captions}
\begin{enumerate}
\item Feynman rules for the $ZZS_a$ and $ZS_aP_b$
couplings ($a=1,2,3;b=1,2$) in the NMSSM.
\item As a function of $m_{S_1}$, 95\% c.l.
upper limits on $\xi^2$
from \cite{alephmssm} (solid lines)
compared to the range in the
NMSSM  (dashed line).
Curve (A) applies, if $S_1$ decays
like a SM
Higgs, curve (B), if it decays invisibly.
If $S_1$ decays into two pseudoscalars,
limit (A) is degraded by less than 10 \%.
The corresponding mass bounds are indicated.
\item The excluded parameter space in the
$M$--$x$ plane
from total $Z$ width
measurements (bright shaded) and direct
neutralino search (dark shaded).
\item The excluded parameter space in the
$A_\lambda - A_k$ plane
(shaded in (a), (c), (e))
and the allowed Higgs mass spectrum
(shaded in (b), (d), (f)) for various parameters.
In (a), (c), (e),
the solid lines denote the contour lines for the mass of the lightest scalar
Higgs, the dotted lines for the mass of the light pseudoscalar Higgs.
The parameter region beyond the $m_{S_1}=0$ GeV contour line and the dashed
line is theoretically excluded as explained in the text.
In (b), (d), (f) the solid line encloses the theoretically allowed domain.
\item The excluded parameter space in the
$A_\lambda - A_k$ plane
(shaded in (a),(c)) and the allowed Higgs mass spectrum
(shaded in (b),(d)) for various parameters.
In (a),(c) the solid lines denote the contour lines for the mass of the
lightest
scalar Higgs, the dotted lines for the mass of the light pseudoscalar Higgs.
The parameter region beyond the $m_{S_1}=0$ GeV contour line and the dashed
line is theoretically excluded as explained in the text.
In (b),(d) the solid line encloses the theoretically allowed domain.
\item Lower bound on the mass of the scalar Higgs
boson as a function
of the coupling parameter $\lambda$ for the singlet vacuum
expectation values $x=1000$ GeV (solid) and $x=100$ GeV (dashed).
\end{enumerate}
\end{document}